\documentclass[%
reprint,
 amsmath,amssymb,
 aps,
]{revtex4-1}
\usepackage{graphicx}
\usepackage{braket}
\usepackage{bm}
\usepackage{amsmath}
\begin{document}
\title{Structure and decay pattern of linear-chain state in $^{14}$C}
\author{T. Baba and M. Kimura}
\affiliation{Department of Physics, Hokkaido University, 060-0810 Sapporo, Japan}
\date{\today}

\begin{abstract}
 The linear-chain states of $^{14}$C are theoretically investigated by using the antisymmetrized
 molecular dynamics. The calculated excitation energies and the $\alpha$ decay widths of the
 linear-chain states were compared  with the observed data reported by the recent
 experiments. The properties of the positive-parity linear-chain states reasonably  agree
 with the observation, that convinces us of the linear-chain formation in the  positive-parity
 states. On the other hand, in the negative-parity states, it is found that the linear-chain
 configuration is fragmented into many states and do not form a single rotational band. 
 As a further evidence of the linear-chain formation, we focus on the $\alpha$ decay pattern. It
 is shown that the linear-chain states decay to the excited states of daughter nucleus $^{10}{\rm
 Be}$  as well as to the ground state, while other cluster states dominantly decay into the ground
 state. Hence, we regard that this characteristic decay pattern is a  strong signature of the
 linear-chain formation.   
\end{abstract}

\maketitle
\section{introduction}
A variety of the $\alpha$ cluster structure are known to exist in light stable nuclei. The most
famous example is the Hoyle state (the $0^+_2$ state of $^{12}{\rm C}$) whose dilute gas-like
3$\alpha$ cluster structure has been studied in detail \cite{uega77,kami81,desc87,enyo97,tohs01,funa03,neff07,funa15} and
identified well today. The linear-chain configuration of 3$\alpha$ clusters in which 
$\alpha$ particles are linearly aligned is an other example of famous and exotic cluster
structure. It  was firstly suggested by Morinaga \cite{mori56} to explain the structure of the
Hoyle state. However, as mentioned above, it turned out that the Hoyle state does not have the
linear-chain configuration but has the dilute gas-like nature. In addition to this, the
instability of the linear-chain  configuration against the bending motion (deviation from the
linear alignment) was pointed out by the antisymmetrized molecular dynamics (AMD) \cite{enyo97}
and Fermionic molecular dynamics (FMD) calculations \cite{neff07}. Thus, the formation of
perfectly linear-aligned configuration in $^{12}{\rm C}^*$ looks negative despite of the many
years of research.  

The interest in the linear-chain configuration is reinforced by the unstable nuclear physics,
because the valence neutrons may stabilize it by their glue-like role. Such glue-like role of
valence neutron is well known for Be isotopes in which 2$\alpha$ cluster core is assisted by
the valence neutrons occupying the molecular-orbits
\cite{seya81,oert96,itag00,yeny99,amd1,amd2}. As a natural consequence, we expect that the
linear-chain configuration can be realized in neutron-rich C isotopes, and this expectation have
been motivating many theoretical and experimental studies
\cite{itag01,gree02,oert03,bohl03,ashw04,oert04,itag06,pric07,suha10,maru10,furu11,baba14,zhao15}. Recently, rather promising candidates of linear-chain
configuration in $^{14}{\rm C}$ were independently reported by two groups
\cite{free14,frit16}. Both groups observed the  $^{4}{\rm He}+{}^{10}{\rm Be}$ resonances above
the $\alpha$ threshold energy in both of positive- and negative-parity. The reported energies of
the positive-parity resonances measured from the $\alpha$ threshold are in reasonable agreement
with the excitation energies of the linear-chain states predicted by Suhara {\it et al.} \cite{suha10} on the
basis of the antisymmetrized molecular dynamics (AMD) calculation. Thus, rather promising evidence
of the linear-chain formation has been found. 

However, there are still several gaps between theory and experiment which must be resolved to
confirm the linear-chain formation in $^{14}{\rm C}$. First, when measured from the ground state
energy, theoretically predicted and experimentally observed excitation energies of the
positive-parity resonances disagree. This may be because the effective interaction used in the
calculation \cite{volk65} do not reproduce the $\alpha$ threshold energy. Second, the experiments
report the negative-parity resonances, while the negative-parity linear-chain states were not
clearly identified in  Ref. \cite{suha10}. Finally, the experiment \cite{free14} reported the
$\alpha$ decay width of the  resonances which is a strong evidence of the $\alpha$ clustering and
must be verified by the theoretical calculation. Thus, further theoretical studies are in need to
identify the linear-chain states in $^{14}{\rm C}$. 

For this purpose, we investigated the linear-chain states in $^{14}{\rm C}$. For
the sake of the quantitative comparison of the excitation energy, we performed AMD calculation
employing Gogny D1S effective interaction \cite{gogn91} which reproduces threshold energies in
$^{14}{\rm C}$. From the AMD wave function, we estimated the $\alpha$ decay widths of the
linear-chain states as well as those of other cluster and non-cluster states. It is found that the
calculated excitation energies of the positive-parity linear-chain states are in good agreement
with the observation, and only the linear-chain states have large $\alpha$ decay widths comparable
with the observed data. Hence, we consider that the linear-chain formation in the positive-parity
state is rather plausible. On the other hand, in the negative-parity state, the linear-chain
configuration is fragmented into many states and do not form a single rotational band.   As a
further evidence of the linear-chain formation, we focus on the $\alpha$ decay pattern. It is
shown that the linear-chain states decay to the excited states of $^{10}$Be as well  as
the $^{10}{\rm Be}$ ground  state, while other cluster states dominantly decay to the  
$^{10}{\rm Be}$ ground state. This characteristic decay pattern is, if it is observed, a strong 
signature of the linear-chain formation.  

The contents of this paper are as follows. In the next section, the AMD framework and the method
to estimate the alpha decay width are briefly outlined. In the III section, the results of the
energy variation and generator coordinate method are presented. In the section IV, the energies
and $\alpha$ width of the linear-chain states are compared with the observed data. In the last
section, we summarize this study.

\section{theoretical framework}

\subsection{variational calculation and generator coordinate method}
The microscopic $A$-body Hamiltonian used in this study reads,
\begin{align}
 \hat{H} = \sum_{i=1}^A \hat{t}_i + \sum_{i<j}^A \hat{v}^N_{ij} + \sum_{i<j}^Z \hat{v}^C_{ij}
 - \hat{t}(c.m.),
\end{align}
where the Gogny D1S interaction \cite{gogn91} is used as an effective nucleon-nucleon interaction
$\hat{v}^N$. It is shown that the Gogny D1S interaction reasonably describes the one neutron-,
$\alpha$- and $^{6}{\rm He}$-threshold energies within 1 MeV error. The Coulomb interaction
$\hat{v}^C$ is approximated by a sum of seven Gaussians. The center-of-mass kinetic energy
$\hat{t}(c.m.)$ is exactly removed.  

The intrinsic wave function $\Phi_{int}$ of AMD is represented by a Slater determinant of single
particle wave packets, 
\begin{align}
 \Phi_{int} ={\mathcal A} \{\varphi_1,\varphi_2,...,\varphi_A \}
 =\frac{1}{\sqrt{A!}}\mathrm{det}[\varphi_{i}({\bm r}_j)],
  \label{EQ_INTRINSIC_WF}  
\end{align}
where $\varphi_i$ is the single particle wave packet which is a direct product of the deformed
Gaussian spatial part \cite{kimu04}, spin ($\chi_i$) and isospin ($\xi_i$) parts,  
\begin{align}
 \varphi_i({\bm r}) &= \phi_i({\bm r})\otimes \chi_i\otimes \xi_i, \label{eq:singlewf}\\
 \phi_i({\bm r}) &= \exp\biggl\{-\sum_{\sigma=x,y,z}\nu_\sigma\Bigl(r_\sigma -\frac{Z_{i\sigma}}{\sqrt{\nu_\sigma}}\Bigr)^2\biggr\}, \\
 \chi_i &= a_i\chi_\uparrow + b_i\chi_\downarrow,\quad
 \xi_i = {\rm proton} \quad {\rm or} \quad {\rm neutron}.\nonumber
\end{align}
The centroids of the Gaussian wave packets $\bm Z_i$, the direction of nucleon spin $a_i, b_i$,
and the width parameter of the deformed Gaussian $\nu_\sigma$ are the variational parameters.  
The intrinsic wave function is projected to the eigenstate of the parity to investigate both of
the positive- and negative-parity states,
\begin{align}
 \Phi^\pi &= P^\pi\Phi_{int}=\frac{1+\pi P_x}{2}\Phi_{int}, \quad \pi=\pm, 
\end{align}
where $P^\pi$ and $P_x$ denote parity projector and operator. Using this wave function, the
variational energy is defined as 
\begin{align}
 E^\pi = \frac{\braket{\Phi^\pi|H|\Phi^\pi}}{\braket{\Phi^\pi|\Phi^\pi}} 
\end{align}
By the frictional cooling method \cite{enyo95}, the variational parameters are determined so that
$E^\pi$ is minimized. In this study, we add the constraint potential to the variational energy,
\begin{align}
 E'^\pi = \frac{\braket{\Phi^\pi|H|\Phi^\pi}}{\braket{\Phi^\pi|\Phi^\pi}} 
 + v_\beta(\braket{\beta} - \beta_0)^2  + v_\gamma (\braket{\gamma} - \gamma_0)^2,
\end{align}
where $\braket{\beta}$ and $\braket{\gamma}$ are the quadrupole deformation parameters of the
intrinsic wave function defined in Ref. \cite{suha10,kimu12}, and $v_\beta$ and $v_\beta$ are chosen large
enough that $\braket{\beta}$ and $\braket{\gamma}$ are equal to $\beta$ and $\gamma$ after the
variation. By minimizing $E'^\pi$, we obtain the optimized wave function 
$\Phi^\pi(\beta,\gamma)=P^\pi\Phi_{int}(\beta,\gamma)$ 
which has the minimum energy for each set of $\beta$ and $\gamma$.

After the variational calculation, the eigenstate of the total angular momentum $J$ is projected
out from  $\Phi^\pi(\beta,\gamma)$,
\begin{align}
 \Phi^{J^\pi}_{MK}(\beta,\gamma) &=  P^{J}_{MK}\Phi^\pi(\beta,\gamma)\nonumber\\
  &=\frac{2J+1}{8\pi^2}
  \int d\Omega D^{J*}_{MK}(\Omega)\hat{R}(\Omega)\Phi^{\pi}(\beta,\gamma).
\end{align} 
Here, $P^{J}_{MK}$, $D^{J}_{MK}(\Omega)$  and $\hat{R}(\Omega)$ are the angular momentum
projector, the Wigner $D$ function and the rotation operator, respectively. The integrals over
Euler angles $\Omega$ are evaluated numerically. 

Then, we perform the GCM calculation by employing the quadrupole deformation parameters $\beta$
and $\gamma$ as the generator coordinate.  The wave function of GCM reads,

\begin{align}
 \Psi^{J^\pi}_{Mn} = \sum_i\sum_Kc^{J^\pi}_{Kin}\Phi^{J^\pi}_{MK}(\beta_i,\gamma_i),\label{eq:gcmwf}
\end{align}
where the coefficients $c^{J^\pi}_{Kin}$ and eigenenergies $E^{J^\pi}_n$ are obtained by solving the
Hill-Wheeler equation \cite{hill54}, 
\begin{align}
 \sum_{i'K'}{H^{J^\pi}_{KiK'i'}c^{J}_{K'i'n}} &= 
 E^{J^\pi}_n \sum_{i'K'}{N^{J^\pi}_{KiK'i'}c^{J^\pi}_{K'i'n}},\\
  H^{J^\pi}_{KiK'i'} &= \braket{\Phi^{J^\pi}_{MK}(\beta_i,\gamma_i)|\hat{H}|
 \Phi^{J^\pi}_{MK'}(\beta_{i'},\gamma_{i'})}, \nonumber\\
  N^{J^\pi}_{KiK'i'} &= \braket{\Phi^{J^\pi}_{MK}(\beta_i,\gamma_i)|
 \Phi^{J^\pi}_{MK'}(\beta_{i'},\gamma_{i'})}.\nonumber
\end{align}
We also calculate the overlap between $\Psi_{Mn}^{J^\pi}$ and the basis wave function of the GCM
$\Phi^{J^\pi}_{MK}(\beta_i,\gamma_i)$, 
\begin{align}
 |\braket{\Phi^{J^\pi}_{MK}(\beta,\gamma)|\Psi^{J^\pi}_{Mn}}|^2/
 \braket{\Phi^{J^\pi}_{MK}(\beta,\gamma)|\Phi^{J^\pi}_{MK}(\beta,\gamma)},
\end{align}
to discuss the dominant configuration in each state described by $\Psi_{Mn}^{J^\pi}$.

\subsection{single particle orbits}
Nucleon single-particle energy and orbit are useful to investigate the motion of the valence
neutrons around the core nucleus. For this purpose, we construct single-particle Hamiltonian
and calculate the neutron single-particle orbits in the intrinsic wave function 
$\Phi_{int}(\beta,\gamma)$.  We first transform the single particle wave packet
$\varphi_i$ to the orthonormalized basis,
\begin{align}
 \widetilde{\varphi}_\alpha = \frac{1}{\sqrt{\lambda_\alpha}}\sum_{i=1}^{A}g_{i\alpha}\varphi_i.  
\end{align}
Here, $\lambda_\alpha$ and $g_{i\alpha}$ are the eigenvalues and eigenvectors of the
overlap matrix $B_{ij}=\langle\varphi_i|\varphi_j\rangle$. Using this basis, the
 single particle Hamiltonian is derived,
\begin{align}
 h_{\alpha\beta} &=
  \langle\widetilde{\varphi}_\alpha|\hat{t}|\widetilde{\varphi}_\beta\rangle + 
  \sum_{\gamma=1}^{A}\langle
  \widetilde{\varphi}_\alpha\widetilde{\varphi}_\gamma|
  {\hat{v}^N+\hat{v}^C}|
  \widetilde{\varphi}_\beta
\widetilde{\varphi}_\gamma -
\widetilde{\varphi}_\gamma\widetilde{\varphi}_\beta\rangle,\nonumber\\ 
 &+\frac{1}{2}\sum_{\gamma,\delta=1}^{A}
 \langle\widetilde{\varphi}_\gamma\widetilde{\varphi}_\delta 
|\widetilde{\varphi}_\alpha^*\widetilde{\varphi}_\beta
\frac{\delta\hat{v}^N}{\delta \rho}|\widetilde{\varphi}_\gamma
\widetilde{\varphi}_\delta - \widetilde{\varphi}_\delta  \widetilde{\varphi}_\gamma
\rangle.
\end{align}
The eigenvalues $\epsilon_s$ and eigenvectors  $f_{\alpha s}$ of $h_{\alpha\beta}$ give the single
particle energies and the single particle orbits,
$\widetilde{\phi}_s = \sum_{\alpha=1}^{A}f_{\alpha s}\widetilde{\varphi}_\alpha$. We also
calculate the amount of the positive-parity component in the single-particle orbit,  
\begin{align}
 p^+ = |\langle \widetilde{\phi}_s|\frac{1+P_x}{2}| \widetilde{\phi}_s\rangle|^2, \label{eq:sp1}
\end{align}
and angular momenta in the intrinsic frame,
\begin{align}
 j(j+1)&= \langle \widetilde{\phi}_s|\hat{j}^2| \widetilde{\phi}_s\rangle, \quad
 |j_z| = \sqrt{\langle \widetilde{\phi}_s|\hat{j}_z^2| \widetilde{\phi}_s\rangle},\label{eq:sp2}\\
 l(l+1)&= \langle \widetilde{\phi}_s|\hat{l}^2| \widetilde{\phi}_s\rangle, \quad
 |l_z| = \sqrt{\langle \widetilde{\phi}_s|\hat{l}_z^2| \widetilde{\phi}_s\rangle},\label{eq:sp3}
\end{align}
which are used to discuss the properties of the single particle orbits.
\subsection{$\bm \alpha$ reduced width amplitude and decay width}
From the GCM wave function, we estimate the $\alpha$ reduced width amplitude (RWA)
$y_{lj^{\pi\prime}}(r)$ which is defined as
\begin{align}
 y_{lj^{\pi\prime}}(r) = \sqrt{\frac{A!}{4!(A-4)!}}
 \langle \phi_\alpha[\phi_{\rm Be}(j^{\pi^\prime})Y_{l0}({\hat r})]_{J^\pi M}
 |\Psi^{J^\pi}_{Mn}\rangle,\label{eq:rwa}
\end{align}
where $\phi_\alpha$ and $\phi_{\rm Be}(j^{\pi'})$ denote the wave functions for $\alpha$ particle
and daughter nucleus $^{10}{\rm Be}$ with spin-parity $j^{\pi'}$. The square of the RWA 
$|ry_{lj^{\pi'}}(r)|^2$ is equal to the probability to observe $\alpha$ particle and 
$^{10}{\rm Be}$ with spin-parity $j^{\pi'}$ at inter-cluster distance $r$ with relative orbital
angular momentum $l$ in the $^{14}{\rm C}$ described by GCM wave function $\Psi_{Mn}^{J^\pi}$. 

Using RWA, the partial $\alpha$ decay widths for the decay process 
$^{14}{\rm C}(J^\pi)\rightarrow \alpha + {}^{10}{\rm Be}(j^{\pi'})$ is estimated as
\begin{align}
 \Gamma_{lj^{\pi'}}^\alpha &= 2P_l(a)\gamma^2_{lj^{\pi'}}(a), \quad
 P_l(a) = \frac{ka}{F^2_l(ka)+G^2_l(ka)}, 
\end{align}
where $a$ denote the channel radius, and the penetration factor $P_l$ is given by the Coulomb
regular and irregular wave functions $F_l$ and $G_l$. The wave number $k$ is given by the decay
$Q$-value and the reduced mass as $k=\sqrt{2\mu E_Q}$. The reduced width $\gamma_{lj^{\pi'}}$ is
\begin{align}
 \gamma^2_{lj^{\pi'}}(a) = \frac{\hbar^2}{2\mu a}[ay_{lj^{\pi'}}(a)]^2.
\end{align}

To calculate RWA with reduced computational cost, we employ the method given in
Ref. \cite{enyoRWA} which suggests an approximation validated for sufficiently large inter-cluster
distance $a$,  
\begin{align}
 |ay_{lj^{\pi'}}(a)|^2 &\simeq \sqrt{\frac{\gamma}{2\pi}}
 |\braket{\Phi^{J\pi}_{BB}(a)| \Psi^{J^\pi}_{Mn}}|^2,\\
 \gamma &= \frac{4(A-4)}{A}\nu_{BB},\nonumber
\end{align}
which means that RWA is reasonably approximated by the overlap between the GCM wave function and
the Brink-Bloch wave function $\Phi_{BB}^{J^\pi}(a)$ composed of the $^{10}{\rm Be}$ and $\alpha$
particle with the Gaussian width parameter $\nu_{BB}$. 

In the case of the present study, since $^{10}{\rm Be}$ is deformed, it must be projected to the
eigenstate of the angular momentum. Therefore, we constructed $\Phi_{BB}^{J^\pi}$ as follows. We
first calculate the intrinsic wave function $\psi_{\rm Be}$ for $^{10}{\rm Be}$ by the AMD energy
variation in which the width parameter $\nu_{BB}$ is fixed to 0.16 fm$^{-2}$. We obtained two
different configurations for $^{10}{\rm Be}$ so-called $\pi$ and $\sigma$ configurations. It is
known that the former is dominant in the ground band and the latter is dominant in the excited
band. Then the $^{10}{\rm Be}$ wave function is projected to the eigenstate of the angular
momentum, and combined with the $\alpha$ cluster wave function to constitute the intrinsic wave
function of $^{14}{\rm C}^*$
\begin{align}
 |\Phi_{BB}(a)\rangle &= \biggl|{\mathcal A} 
 \biggl\{\psi_\alpha\left(-\frac{10}{14}a\right)
 \hat{P}^{j}_{m0}\psi_{\rm Be}\left(\frac{4}{14}a\right)\biggl\}\biggl\rangle, \label{eq:bbwf} 
\end{align}
where we assumed that $\psi_{\rm Be}$ is parity and axially symmetric, that is validated by 
the numerical check. Then, the reference wave function is constructed by the angular momentum
projection of the total system. 
\begin{align}
 \Phi^{J\pi}_{BB}(a) &=
 \frac{1}{N}\frac{2l+1}{2J+1}\sum_mC^{Jm}_{l0jm}\hat{P}^{J^\pi}_{Mm}|\Phi_{BB}(a)\rangle,
 \label{eq:norm}
\end{align}
Here $C^{Jm}_{l0jm}$ and $N$ denotes the Clebsch-Gordan coefficient and the normalization factor. 
The summation over $m$ is needed to project the relative angular momentum between $^{10}{\rm Be}$
and $\alpha$ particle to $l$.

\section{Results}
\subsection{Energy surface and intrinsic structures}
\begin{figure}[h]
 \centering
 \includegraphics[width=1.0\hsize]{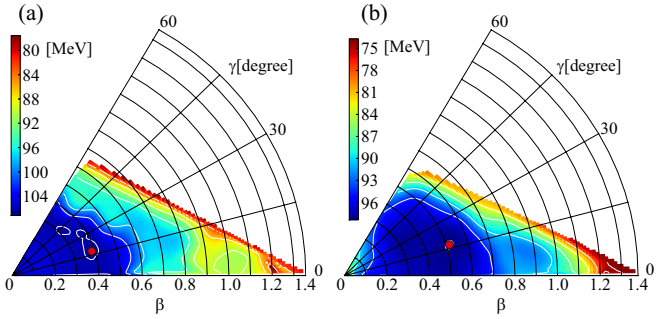}
 \caption{(color online) The angular momentum projected energy surface for (a) the $J^\pi=0^+$
 state and (b) $J^\pi=1^-$ state as functions of quadrupole deformation parameters $\beta$ and
 $\gamma$.  The circles show the position of the energy minima.} 
 \label{fig:surface}
\end{figure}
\begin{figure}[h]
 \centering
 \includegraphics[width=1.0\hsize]{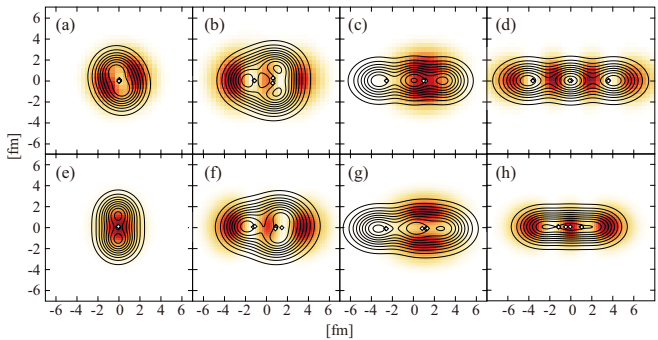}
 \caption{(color online) The density distribution of (a)-(d) the positive states and (e)-(h)
 negative parity states.  The contour lines show the proton density distributions.
 The color plots show the single particle orbits occupied by the most weakly bound neutron.
 Open boxes show the centroids of the Gaussian wave packets describing protons.} 
 \label{fig:density}
\end{figure}
\begin{table}[h]
\caption{The properties of the most weakly bound proton and neutron orbits in the configurations
 shown in Fig. \ref{fig:density} (a)-(h). The column occ. shows the number of the nucleon occupying
 the orbit. When two valence nucleons occupy the almost degenerated orbits, the single-particle
 properties are averaged and occ. is equal to 2. 
 Other columns show the single particle energy $\varepsilon$ in MeV, the amount of the
 positive-parity component $p^+$ and the angular momenta defined by
 Eqs. (\ref{eq:sp1})-(\ref{eq:sp3}).} 
\label{table:spo}
\begin{center}
 \begin{ruledtabular}
  \begin{tabular}{clccccccc} 
   & & occ.& $\varepsilon $ &$p^+$ & $j$ & $|j_{z}|$ & $l$ & $|l_{z}|$ \\ \hline
	(a) &proton  & 2 & -17.4 & 0.00 & 1.5 & 1.5 & 1.1 & 1.0 \\
            &neutron & 2 & -6.6 & 0.22 & 1.1 & 0.6 & 1.2 & 0.9 \\
	(b) &proton  & 2 & -14.1 & 0.08 & 1.6 & 1.5 & 1.2 & 1.0 \\ 
	    &neutron & 2 & -5.3 & 0.98 & 2.2 & 0.5 & 1.8 & 0.3 \\ 
	(c) &proton  & 2 & -12.5 & 0.97 & 2.2 & 0.5 & 2.0 & 0.2 \\
	    &neutron & 2 & -7.0 & 0.09 & 1.8 & 1.5 & 1.4 & 1.0 \\
	(d) &proton  & 2 & -15.6 & 0.99 & 2.5 & 0.5 & 2.3 & 0.1 \\
	    &neutron & 2 & -4.4 & 0.01 & 2.8 & 0.5 & 2.6 & 0.1 \\ \hline
	(e) &proton & 2 & -16.0 & 0.00 & 1.5 & 1.4 & 1.1 & 1.0 \\
	    &neutron & 1 & -3.8 & 0.99 & 2.2 & 0.5 & 1.8 & 0.4 \\
	(f) &proton  & 1 & -12.6 & 0.53 & 1.9 & 0.9 & 1.6 & 0.8 \\ 
            &neutron & 2 & -6.6 & 0.98 & 2.1 & 0.6 & 1.8 & 0.3 \\ 
	(g) &proton  & 1 & -12.4 & 0.72 & 2.3 & 0.9 & 2.1 & 0.6 \\ 
            &neutron & 2 & -7.2 & 0.11 & 1.9 & 1.4 & 1.6 & 1.0 \\ 
	(h) &proton  & 1 & -13.1 & 0.52 & 1.9 & 1.0 & 1.6 & 0.8 \\ 
	    &neutron & 2 & -8.2 & 0.92 & 2.2 & 0.7 & 1.9 & 0.4 \\ 
  \end{tabular}
  \end{ruledtabular}
\end{center}
\end{table}

Figure \ref{fig:surface} (a) shows the energy surface as functions of quadrupole deformation
parameters $\beta$ and $\gamma$ for $J^\pi=0^+$ states obtained by the constraint
variational calculation and angular momentum projection. The circles on the energy surfaces show
the position of the energy minima.

The energy minimum of the $0^+$ state is located at $(\beta,\gamma)=(0.36,14^\circ)$ with the
binding energy of 106.1 MeV. It is interesting that this minimum state is deformed, as seen
in its intrinsic density distribution shown in Fig. \ref{fig:density} (a), despite of the neutron 
magic number $N=8$. However, the deformation is not large enough to break the neutron magicity as
the last valence neutron occupies $p$-wave that can be deduced from the density distribution
of the valence neutron orbit shown in Fig. \ref{fig:density} (a).

In the oblate deformed region, the different structure which we call triangular configuration
appears. Fig. \ref{fig:density} (b) shows the density distribution of the wave function located at 
$(\beta,\gamma)=(0.60,25^\circ)$. The proton density distribution have triangular shape showing
possible formation of 3$\alpha$ cluster core with triangle configuration. Indeed, as mentioned
later, the excited states composed of this configuration have larger $\alpha$ reduced widths than
the ground state. Owing to the parity asymmetric shape, the valence proton orbit is an admixture
of the positive- and negative-parity component as confirmed from the properties of the
single-particle orbit listed in Table \ref{table:spo}. The table also shows that two valence
neutrons occupy positive-parity orbit ($sd$-shell) indicating the $2\hbar\omega$
excitation. It is noted that a similar configuration, {\it i.e.} a triangular 3$\alpha$ cluster
core with $2\hbar\omega$ excited valence neutrons, was also found in $^{16}{\rm C}$ \cite{baba14}.

\begin{figure}[h]
 \centering
 \includegraphics[width=1.0\hsize]{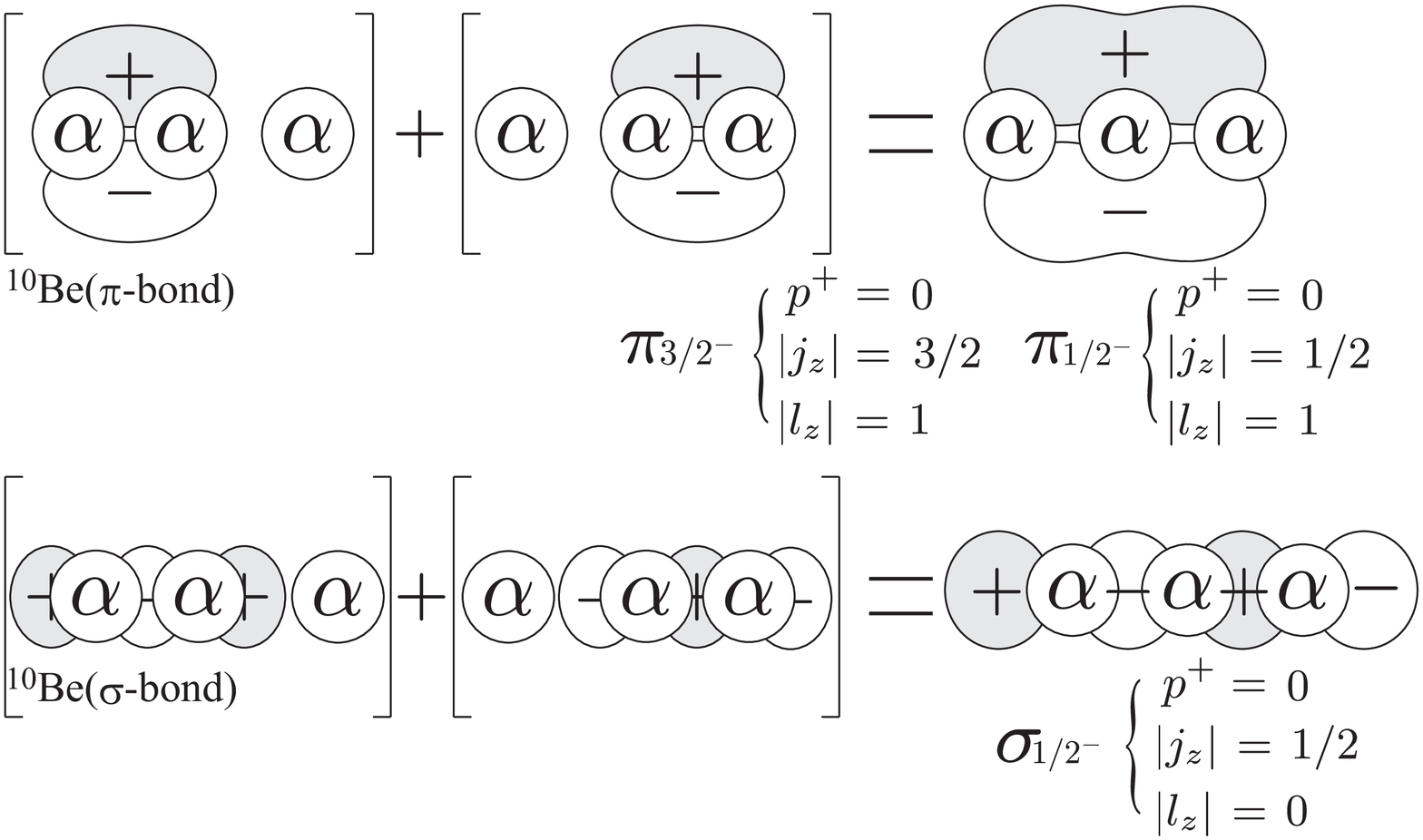}
 \caption{The schematic figure showing the $\pi$ and $\sigma$-orbits around the linear chain. 
 The combination of the $p$ orbits perpendicular to the symmetry axis generates
 $\pi$ orbits, while the combination of parallel orbits generates $\sigma$ orbit.} 
 \label{fig:illust_mol}
\end{figure}

In the strongly deformed region, the linear-chain configurations appear. There is an energy
plateau around the local energy minimum at $(\beta,\gamma)=(1.00,0)$ which approximately
corresponds to the deformation ratio equal to $3:1$. As clearly seen in its density distribution
shown in Fig. \ref{fig:density} (c), this local energy minimum has pronounced 3$\alpha$ cluster
structure with linear alignment. The properties of the proton single-particle orbit shows that the
last two protons are promoted into $sd$-shell, that is because of the Pauli principle in the
linear-chain configuration. The density distribution and properties of the valence neutron orbits
show that they correspond to the $\pi$-orbit of the molecular-orbit picture which is
schematically illustrated in Fig. \ref{fig:illust_mol}. Namely, the valence neutron orbit is a
linear combination of the $p$-orbits perpendicular to the symmetry axis and has the angular
momenta $|j_{z}|=1.5$ and $|l_{z}|=1.0$. We call this configuration $\pi$-bond linear
chain in the following. This property of the valence neutron orbit is common to 
that found in $^{16}{\rm C}$ \cite{baba14}. However, it should be noted that the the $\pi$-orbit of 
$^{14}{\rm C}$ do not have parity symmetric distribution but is localized 
between the center and right $\alpha$ clusters. In other words, this configuration is the parity
asymmetric and has $^4$He$+^{10}$Be-like structure, which is consistent with the discussion made
in Ref. \cite{suha10,suha11}. Because this linear-chain configuration and the triangular
configuration explained above have asymmetric internal structures, we expect that the
corresponding negative-parity partners may exist and constitute the inversion doublets.  

With further increase of the deformation, the other linear-chain configuration which we call
$\sigma$-bond linear chain appears around $(\beta,\gamma)=(1.27,0)$ which was not mentioned in
Ref. \cite{suha10}. From the density distribution (Fig. \ref{fig:density} (d)), it is clear that
this configuration has another valence neutron orbit that correspond to the $\sigma$-orbit which
is a linear combination of $p$-orbit parallel to the symmetry axis and has the angular momenta
$|j_{z}| \approx 0.50$ and  $|l_{z}|\approx 0$. It is interesting to note that this configuration
has parity symmetric shape, and hence, do not have its negative-parity partner. 

The energy minimum of the energy surface for the $1^-$ states (Fig. \ref{fig:surface} (b)) is
located at $(\beta,\gamma)=(0.52,19^\circ)$ with the binding energy of $-98.2$ MeV, and its density
distribution is described in Fig. \ref{fig:density}(e). From Tab. \ref{table:spo}, we see that
this minimum has the same proton configuration with the positive-parity minimum, but a neutron is
excited into $sd$-shell from $p$-shell ($1p1h$ configuration). 

Because the triangular configuration and the $\pi$-bond linear-chain of the positive parity are
parity asymmetric, their counterparts appear in the negative-parity states. Fig. \ref{fig:density}
(f) and (g) show the triangular configuration and the $\pi$-bond linear-chain configuration in the
negative-parity state located at $(\beta,\gamma) = (0.72,14^\circ)$ and  $(1.05, 3^\circ)$,
respectively. Although the cluster cores are more distorted than the positive-parity states, their
neutron single-particle configurations are similar to their positive-parity counterparts. However,
as explained in the next section, these cluster configurations do not form single rotational
band and are fragmented into many states because of the mixing with other non-cluster states. For
example, the negative-parity $\pi$-bond linear-chain strongly mixes with the strongly deformed
non-cluster states such as the configuration shown in Fig. \ref{fig:density} (h), that makes it
impossible to identify the linear-chain rotational state uniquely.

\subsection{Excitation spectrum}
\begin{figure*}[t]
 \includegraphics[width=0.9\hsize]{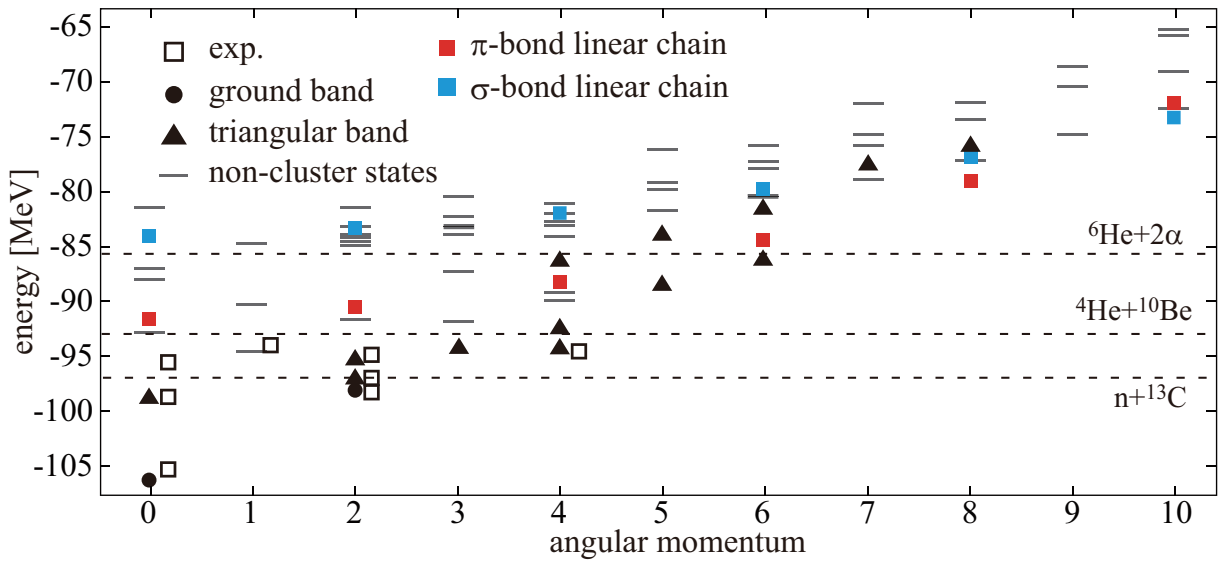}
 \caption{(color online) The positive-parity energy levels up  to $J^\pi=10^+$.  Open
 boxes show the observed states with the definite spin-parity assignments taken from
 Ref. \cite{ajze91}, and other symbols show the calculated result.  The filled circles, triangles
 and filled boxes show the ground, triangular and linear-chain bands, while lines show the 
 non-cluster states.}  \label{fig:spectrum+} 
\end{figure*}

Figure \ref{fig:spectrum+} shows the spectrum of the positive-parity states obtained by the GCM
calculation. The properties of the several selected states are listed in Tab. \ref{tab:band}. We
classified the excited states which have large $\alpha$ reduced widths as cluster states. The
detail of the $\alpha$ reduced widths is given in the section IV B. In the case of the
positive-parity states, the cluster states are assigned in the rotational bands without
uncertainty, because the band member states are connected by the strong $E2$ transitions as listed
in Tab. \ref{table:be2}.
\begin{table}[h]
 \caption{Excitation energies (MeV) and  proton and neutron root-mean-square radii (fm) of several 
 selected states.  Numbers in the  parenthesis are the observed data \cite{ajze91,ange13}.}\label{tab:band}   
\begin{center}
 \begin{ruledtabular}
  \begin{tabular}{lcccc} 
   band & $J^\pi$ & $E_x$ & $r_{p}$ & $r_{n}$\\
   \hline
   ground & $0^+_1$ & 0.00 & 2.53 & 2.58 \\
               & $2^+_1$ & 8.41  & 2.58 & 2.69 \\
   & & (7.01) & (2.34) & \\
   \hline
   triangular & $0^+_2$ & 7.49  & 2.67 & 2.92 \\
   $K^\pi=0^+$     & $2^+_2$ & 9.26  & 2.64 & 2.83 \\
                   & $4^+_1$ & 12.00  & 2.65 & 2.89 \\
   \hline
   triangular & $2^+_3$ & 10.99  & 2.68 & 2.92 \\
   $K^\pi=2^+$     & $3^+_1$ & 12.03 & 2.68 & 2.92 \\
                   & $4^+_2$ & 13.83 & 2.68 & 2.92 \\
   \hline
   $\pi$-bond & $0^+_4$ & 14.64 & 3.27 & 3.20 \\
   linear chain & $2^+_5$ & 15.73  & 3.37 & 3.28 \\
                     & $4^+_5$ & 17.98  & 3.33 & 3.24 \\
                     & $6^+_2$ & 21.80  & 3.39 & 3.30 \\
   \hline
   $\sigma$-bond & $0^+_7$ & 22.16 & 3.91 & 4.12 \\
   linear chain & $2^+_{10}$ & 22.93 & 4.02 & 4.21 \\
                     & $4^+_{11}$ & 24.30 & 3.97 & 4.15 \\
  \end{tabular}
 \end{ruledtabular}
 \end{center}
\end{table}

\begin{table}[h!]
 \caption{The calculated in-band $B(E2)$ strengths for the low-spin positive-parity states in unit
 of  $e^2\rm fm^4$.  For the negative-parity states, the transitions between the low-spin
 fragmented cluster states (diamonds in Fig. \ref{fig:spectrum-}) are shown and the
 transitions less than 10 $e^2\rm fm^4$ are not shown. The number in parenthesis is the observed
 data \cite{be01}. } 
\label{table:be2}
\begin{center}
 \begin{ruledtabular}
 \begin{tabular}{lcc}
   & $J_i\rightarrow J_f$ & $B(E2;J_i\rightarrow J_f)$ \\ \hline
  ground $\rightarrow$ ground& $2^{+}_{1}\rightarrow 0^{+}_{1}$ & 8.1 (3.74)\\ \hline
  triangular $K^\pi=0^+$& $2^{+}_{2} \rightarrow 0^{+}_{2}$ & 7.6 \\
  $\rightarrow$ triangular $K^\pi=0^+$ & $4^{+}_{1}\rightarrow 2^{+}_{2}$ & 7.9 \\
                                        & $6^{+}_{1}\rightarrow 4^{+}_{2}$ & 19.8 \\\hline
  triangular $K^\pi=2^+$ & $3^{+}_{1} \rightarrow 2^{+}_{3}$ & 17.6 \\
  $\rightarrow$ triangular $K^\pi=2^+$ & $4^{+}_{2}\rightarrow 3^{+}_{1}$ & 8.5 \\
                                        & $4^{+}_{2}\rightarrow 2^{+}_{3}$ & 5.4 \\\hline
  $\pi$-bond linear chain & $2^{+}_{5}\rightarrow 0^{+}_{4}$ & 165.5\\
  $\rightarrow$ $\pi$-bond linear chain & $4^{+}_{5}\rightarrow 2^{+}_{5}$ & 257.4\\
                                          & $6^{+}_{2}\rightarrow 4^{+}_{5}$ & 276.5\\ \hline
  $\sigma$-bond linear chain & $2^{+}_{10}\rightarrow 0^{+}_{7}$ & 441.9\\
  $\rightarrow$ $\sigma$-bond linear chain & $4^{+}_{11}\rightarrow 2^{+}_{10}$ & 655.9\\ \hline
  negative parity states     & $3^{-}_{4}\rightarrow 1^{-}_{3}$ & 21.9\\
                             & $3^{-}_{5}\rightarrow 1^{-}_{3}$ & 32.4\\
                             & $3^{-}_{6}\rightarrow 1^{-}_{5}$ & 60.1\\
                             & $3^{-}_{10}\rightarrow 1^{-}_{5}$ & 31.5\\
                             & $5^{-}_{2}\rightarrow 3^{-}_{4}$ & 63.0\\
                             & $5^{-}_{4}\rightarrow 3^{-}_{5}$ & 54.5\\
                             & $5^{-}_{7}\rightarrow 3^{-}_{6}$ & 53.9\\
 \end{tabular}
 \end{ruledtabular}
\end{center}
\end{table}

The ground state and the first excited state ($2^+_1$) are dominantly composed of the
configurations around the energy minimum of the energy surface. The ground state has the largest
overlap with the configuration shown in Fig. \ref{fig:density} (a) that amounts to 0.98, and the
calculated binding energy is $-106.3$ MeV that is reasonably close to the observed value of $-105.3$
MeV. The excitation energy of the $2^+_1$ state is also reasonably described. 

Owing to its triaxial deformed shape, the triangular configuration generates the rotational bands
built on the $0^+_2$ and $2^+_3$ states that are shown by triangles in
Fig. \ref{fig:spectrum+}. We call them  $K^\pi=0^+$ and $2^+$ bands, respectively  in the
following, although the mixing of the $K$ quantum number in their GCM wave functions is not
negligible. Compared to the linear-chain states, these bands have less pronounced clustering and
$\alpha$  clusters are considerably distorted, therefore the band head energies are well below the
cluster thresholds. The member states have large overlap with the configuration shown in
Fig. \ref{fig:density} (b) which amount to, for example, 0.91 in the case of the $0^+_2$ state.   

The linear-chain configurations generate two rotational bands in Fig. \ref{fig:spectrum+}.
The first one which we call $\pi$-bond linear-chain band is built on the $0^+_4$ state at 14.6 MeV 
close to the $\alpha$ threshold energy and composed of the $\pi$-bond linear-chain configurations. 
The band head state (the $0^+_4$ state) has large overlap with the configuration shown in 
Fig.\ref{fig:density} (c) which amounts to 0.87. The other band which we call $\sigma$-bond
linear-chain band is built on the $0^+_7$ state at 22.2 MeV (about 9.18 MeV above the $\alpha$
threshold) and composed of the $\sigma$-bond linear-chain configurations shown in
Fig.\ref{fig:density} (d). This  intrinsic wave function has the largest overlap with the band
head state that amounts to 0.99.  The $\pi$-bond linear-chain band is the candidate of the
observed resonances and the comparison with the observation is discussed in the next section. 

\begin{figure*}[t]
 \includegraphics[width=0.9\hsize]{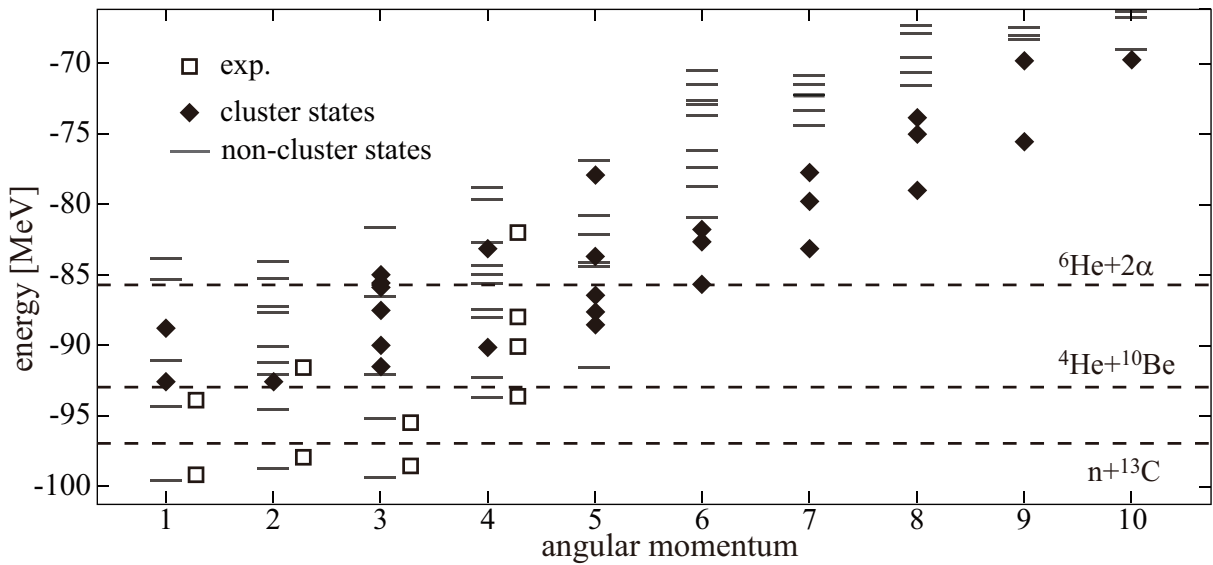}
 \caption{The  negative-parity energy levels up to $J^\pi=10^-$.  Open boxes show the observed
 states with the definite spin-parity assignments taken from Ref. \cite{ajze91}, and other symbols
 show the calculated result.  The diamonds show the cluster states having non-negligible $\alpha$
 reduced widths,  while lines show the non-cluster states.}  \label{fig:spectrum-} 
\end{figure*}

In the case of the negative-parity states shown in Fig. \ref{fig:spectrum-}, it is found that the
$\alpha$ cluster configurations are fragmented into many excited states. As a result, the $E2$
transition strengths are also spread to several states, and it makes the band assignment
ambiguous. For example, as listed in Tab. \ref{table:be2}, there are two $3^-$ states that
strongly decay to the $1^-_3$ state. Those fragments of the cluster configurations are shown by
diamond symbols which are mainly composed of the configurations  shown in
Fig. \ref{fig:density}(f), (g), (h) and other non-cluster configurations. For example, the $1^-_3$
state has the largest overlap with the configuration shown in Fig. \ref{fig:density} (h) that
amounts to 0.93. But, at the same time, this state also have large overlap with the triangular
configuration shown in Fig. \ref{fig:density} (f) and the $\pi$-bond linear-chain shown in
Fig. \ref{fig:density} (g) which amounts to 0.85 and 0.70, respectively. This means that these
fragmented states are the mixture of cluster states and non-cluster states. The fragmentation of
the cluster configurations can be more clearly seen in their $\alpha$ reduced widths which are
discussed in the next section.

\section{Discussion}
 \subsection{Excitation energies of the linear-chain bands}
\begin{figure}[t]
 \includegraphics[width=\hsize]{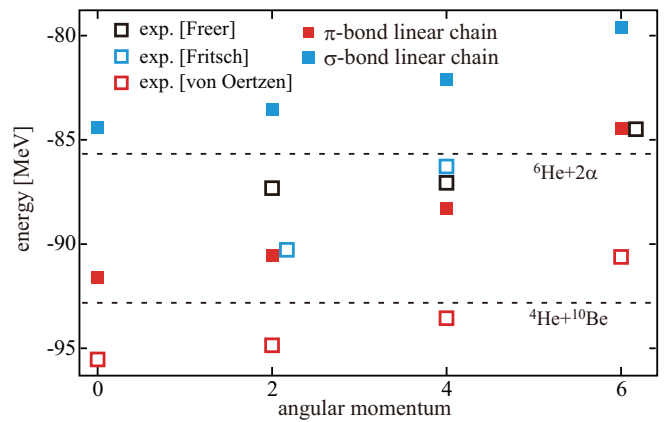}
 \caption{(color online) The calculated and observed linear-chain candidates in positive
 parity. Open boxes show the  observed data reported by Refs. \cite{oert04,free14,frit16}. Filled boxes show the
 energies of the $\pi$-bond and  $\sigma$-bond linear chain states.}  \label{fig:spectrum+2} 
\end{figure}
In this section, we focus on the excitation energies of the linear-chain bands and compare them
with the experiments \cite{oert04,free14,frit16}. The results of the present calculation and the
experimental candidates are summarized in Fig. \ref{fig:spectrum+2}. By the measurement of the 
$^9$Be($^7$Li,d)$^{14}$C reaction, von Oertzen {\it et al.} \cite{oert04} reported a candidate of the
linear-chain band whose band-head energy is below the $\alpha$ threshold energy. Freer {\it et
al.} \cite{free14} and  Fritsch {\it et al.} \cite{frit16} independently 
measured $^{4}{\rm He}+{}^{10}{\rm Be}$ resonant scattering using radioactive ${}^{10}{\rm Be}$
beam, and reported the candidates above the threshold energy. The resonance energies of $4^+$
state reported by Freer {\it et al} and Fritsch {\it et al.} are close to each other, but those of 
$2^+$ state differ. However, it must be kept in mind that the assignment of the $2^+$ state by
Freer {\it et al.} is tentative as mentioned in their report.

Then, we see that the calculated energy of the $\pi$-bond linear chain is close to  the
resonances observed in the $^{4}{\rm He}+{}^{10}{\rm Be}$ resonant scattering except for the
tentatively assigned $2^+$ state. In addition, as discussed in the next section, the $\alpha$
reduced widths of the $\pi$-bond linear chain and those observed resonances are close to each
other. Hence, we conclude that the resonances observed in the $^{4}{\rm He}+{}^{10}{\rm Be}$
resonant scattering should be the $\pi$-bond linear chain. The excited states reported by von
Oertzen {\it et al} are approximately 5 MeV lower than the $\pi$-bond linear chain, and it
energetically corresponds to the triangular band. The measurement of the $\alpha$ widths of those
candidate will make this assignment sure. The $\sigma$-bond linear chain is energetically located
higher than any observed resonances and does not have the experimental counterpart. As we 
see later, the $\sigma$-bond linear chain is dominantly composed of the 
$^{4}{\rm He}+{}^{10}{\rm Be}(0^+_2)$ and $^{4}{\rm He}+{}^{10}{\rm Be}(2^+_2)$
component. Therefore, we consider that it is not easy to populate this band by ordinary transfer
reaction or resonant scattering. 

\begin{figure}[t]
 \includegraphics[width=\hsize]{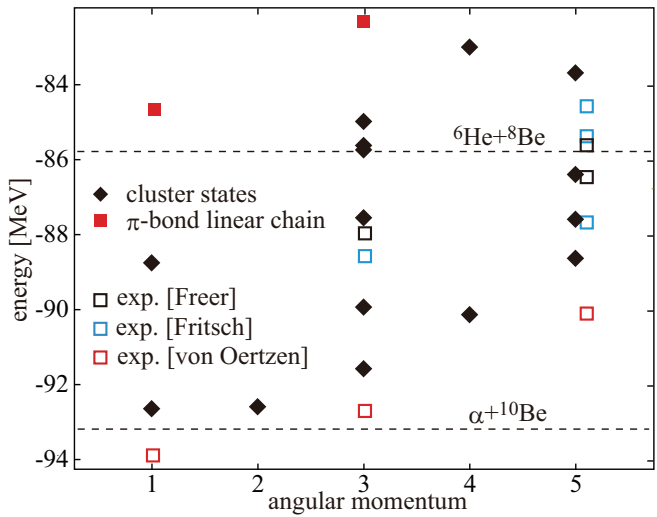}
 \caption{(color online) The calculated cluster states and observed linear-chain candidates in
 negative parity. Open boxes show the  observed data reported by Refs. \cite{oert04,free14,frit16}. Filled diamonds
 are the fragmented excited states with cluster configuration. Red boxes shows the $\pi$-bond
 linear-chain projected to negative-parity.}  \label{fig:spectrum-2} 
\end{figure}
Fig. \ref{fig:spectrum-2} summarizes the negative-parity results. In contrast to the positive
parity, there are so many fragments of the cluster configurations in the theoretical result. As a
result, the correspondence between the theory and experiment is not unique. We also performed an
additional test calculation. We pickup the $\pi$-bond linear-chain configuration with positive
parity  shown in Fig. \ref{fig:density} (c) and artificially project it to the negative-parity to
estimate the energy of the ideal $\pi$-bond linear chain with negative-parity. The results is
shown by the red filled boxes in Fig. \ref{fig:spectrum-2}. We see that the energy of the ideal
linear-chain is too high to be assigned to the observed resonances. Thus, the present calculation
does not support the formation of the linear chain in negative parity.

\subsection{Reduced width}
\begin{figure*}[h]
 \centering
 \includegraphics[width=0.8\hsize]{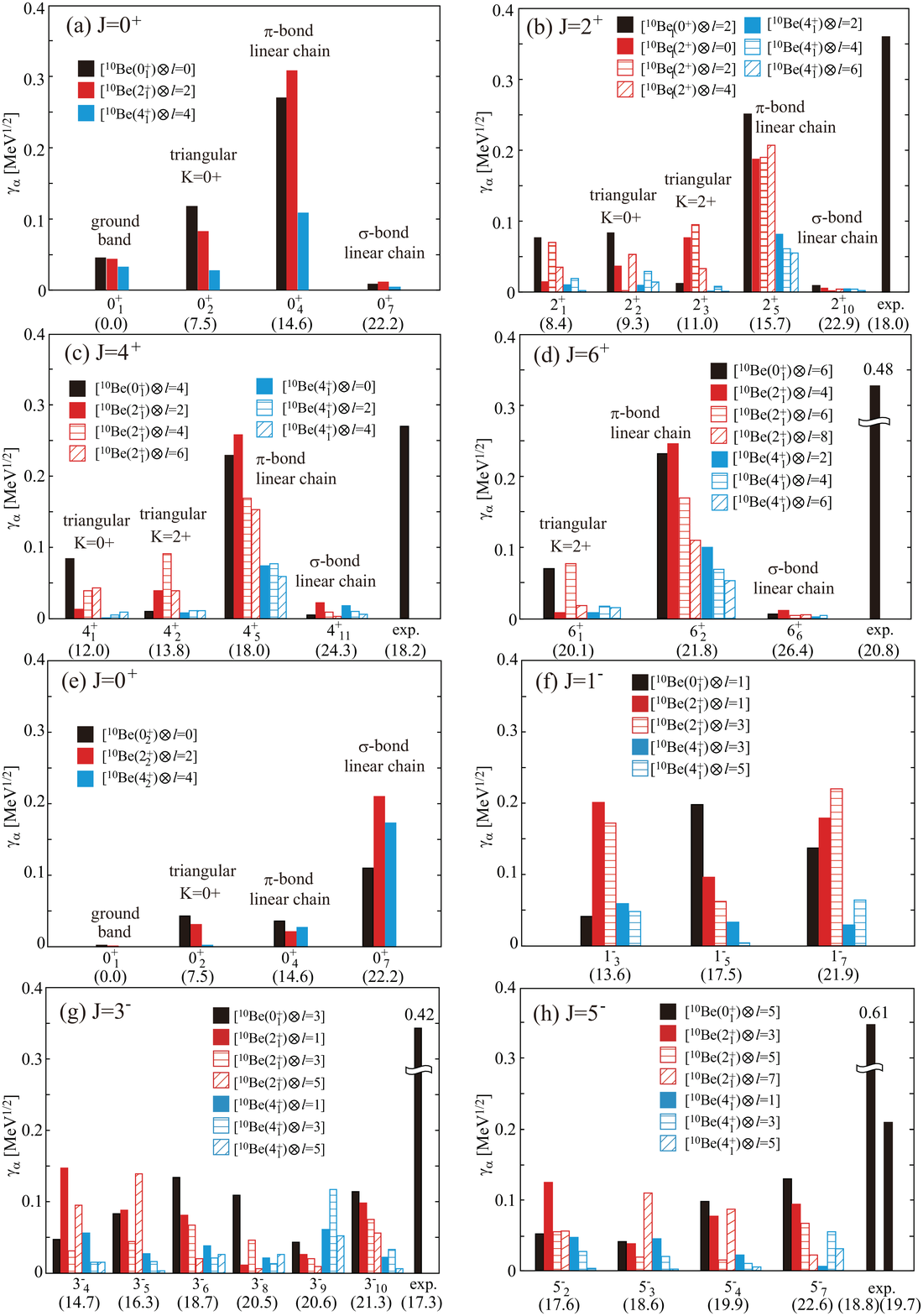}
 \caption{(color online) The calculated $\alpha$-decay reduced widths compared with the observed 
 widths reported in Ref. \cite{free14}. Panels (a)-(d) show the decay of the  positive-parity states
 to the ground band of $^{10}{\rm Be}$ ($\pi$-bonded $^{10}{\rm Be}$).  Panel (e) shows the decay
 of the $0^+$ states to the excited band of $^{10}{\rm Be}$ ($\sigma$-bonded $^{10}{\rm Be}$). 
 Panels (f)-(h) show the decay of the negative-parity states  to the ground band of 
 $^{10}{\rm Be}$ ($\pi$-bonded $^{10}{\rm Be}$). Numbers in parenthesis show the excitation
 energy.}  
 \label{fig:rwa+} 
\end{figure*}

Figure \ref{fig:rwa+} shows the $\alpha$ reduced widths of 
several selected low-spin states with positive parity. The decay channels are indicated as
$[^{10}{\rm Be}(j^\pi)\otimes l]$ where $j^\pi$ and $l$ denote the angular momenta of the 
$^{10}{\rm Be}$ ground band and the relative motion between $^{10}{\rm Be}$ and $\alpha$ particle, 
respectively. Here, $^{10}$Be is assumed to have two neutrons in $\pi$-orbit. The dominance of
the $\pi$-bond configuration in the $^{10}{\rm Be}$ was confirmed by the observations \cite{powe70}.

There are two prominent features to be noted in these results. The first is the magnitude of the
reduced widths. The $\pi$-bond linear-chain band (the $0^+_4$, $2^+_5$, $4^+_5$ and $6^+_2$
states) have large reduced widths compared to the triangular bands and the ground state. It is
also noted that the $\alpha$ reduced widths of other excited states are also smaller than the
$\pi$-bond linear-chain band, and even smaller than or comparable with the triangular
bands. Hence, in the  calculated energy region, the $\pi$-bond linear chain band has the largest
reduced width. In Fig. \ref{fig:rwa+}(b)-(d), the observed reduced widths of the linear-chain candidates
\cite{free14} are also shown for $2^+$, $4^+$ and $6^+$ states. Since the decay to the $^{10}{\rm
Be}$ ground state was {\it assumed} in the R-matrix analysis made in Ref. \cite{free14}, those values
may be compared with the calculated results for the $[^{10}{\rm Be}(0^+_1)\otimes l]$ channel, and 
we see that only the $\pi$-bond linear-chain band can explain the magnitude of the observed
reduced widths. Thus, both of the observed excitation energies and reduced widths are
reasonably  explained by the $\pi$-bond linear-chain band, and we consider that the linear-chain 
formation in the positive-parity looks plausible. 

It is also interesting to note that the other linear-chain band, {\it i.e.} the $\sigma$-bond
linear-chain band, has suppressed reduced widths despite of their prominent $\alpha$
clustering. The reason is simple. Because the $\sigma$-bond linear-chain band does 
not have valence neutron in $\pi$-orbit, it is orthogonal to the decay channels to the
$^{10}{\rm Be}$ ground state that has $\pi$-orbit neutron. This is confirmed in
Fig. \ref{fig:rwa+} (e) where the reduced widths for the decays to the $^{10}{\rm Be}$ with
$\sigma$-bond (the excited rotational band of $^{10}{\rm Be}$) are shown. Since other bands do
not have valence neutrons in $\sigma$-orbit, their reduced widths are suppressed, and only the
$\sigma$-bond linear-chain band has large widths. 

Another point to be noted is the decay pattern of the $\pi$-bond linear-chain band. The reduced
widths in the [$^{10}$Be$(2^+_1)\otimes l$] channels are as large as or even larger than those in
the [$^{10}$Be$(0^+_1)\otimes l$] channel. This dominance of the $^{10}{\rm Be}(2^+_1)$ 
component in the $\pi$-bond linear-chain band is owe to its unique structure. When three $\alpha$
particles are linearly aligned, because of the strong angular correlation between $\alpha$
particles, the $^{10}{\rm Be}(2^+_1)$ and $^{10}{\rm Be}(4^+_1)$ components become large. This
property is in contrast to the Hoyle state where $\alpha$ particles are mutually orbiting with
$l=0$, and hence, the $^{8}{\rm Be}(0^+_1)$ component dominates \cite{funa15}. Similar property of the
linear-chain  configuration were also discussed  in $^{12}{\rm C}$\cite{suzu72}. Therefore, if the
large contamination of the $^{10}{\rm Be}(2^+_1)$ component is confirmed, it will be a strong
evidence for the linear-chain formation.

\begin{figure}[h]
 \centering
 \includegraphics[width=0.7\hsize]{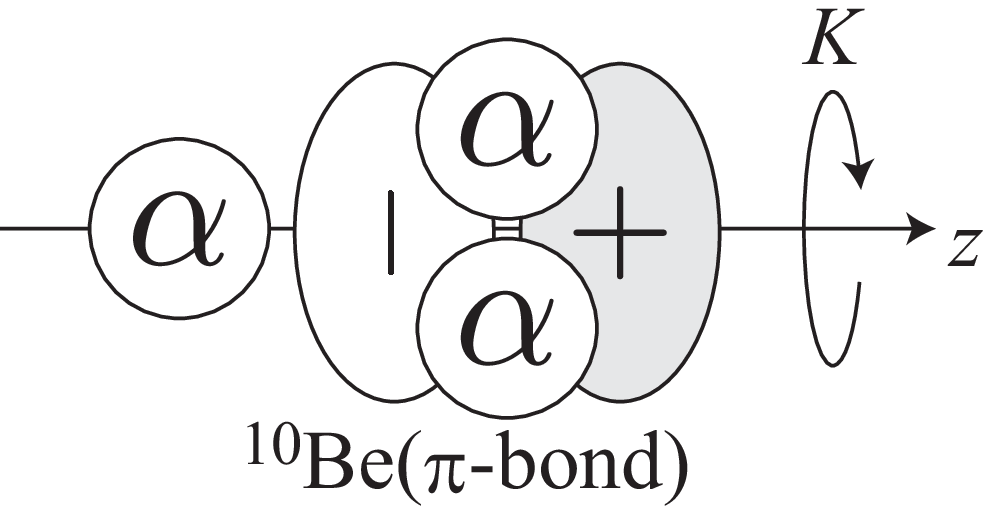}
 \caption{The schematic figure which explain the relationship between the $K$ quantum number and 
 the angular momentum of $^{10}{\rm Be}$. } 
 \label{fig:tri_illust}
\end{figure}

The $^{10}{\rm Be}(2^+_1)$ component in the triangular bands also show an interesting
feature. There are two triangular bands with $K^\pi=0^+$ and $K^\pi=2^+$. The $0^+_2$, $2^+_2$ and
$4^+_1$ states are the member of the $K^\pi=0^+$ band, while the $2^+_3$, $4^+_2$ and $6^+_1$ are
member of the $K^\pi=2^+$ band. Here, we clearly see that the $^{10}{\rm Be}(0^+_1)$ component is 
dominant in the $K^\pi=0^+$ band, while the $^{10}{\rm Be}(2^+_1)$ component is dominant in the
$K^\pi=2^+$ band. This feature is explained by Fig. \ref{fig:tri_illust}. In the triangular bands,
$^{4}$He and $^{10}$Be are placed in a triangular shape and the intrinsic $z$-axis is chosen to be
perpendicular to the deformation axis of $^{10}{\rm Be}$. Since the $K$ quantum number is the
angular momentum directed to the intrinsic $z$-axis, $K$ must be equal to the angular momentum of
$^{10}{\rm Be}$.  This makes the difference in the amount of the $^{10}{\rm Be}(2^+_1)$ component
in the $K^\pi=0^+$ and $2^+$ bands. 

For the negative parity, we show the states which have the reduced widths larger than 0.1
MeV$^{1/2}$ in Fig. \ref{fig:rwa+} (f)-(h). We can see that there are many states which have
non-negligible $\alpha$ reduced widths, and not able to identify the linear-chain band. As already
mentioned, the linear-chain configurations are coupled with the non-cluster configurations and
fragmented into many states as found in Ref. \cite{suha10}.  We also see that none of the calculated
state can explain the observed reduced widths that are twice larger than the present results. This
requires further theoretical study of the negative-parity states, although the current result
looks negative to the linear-chain formation in the negative-parity.

\section{SUMMARY}
In order to investigate the existence of the linear-chain state, we have studied the excited
states of $^{14}$C based on the AMD calculations.  

In the positive-parity states, we found that three different configurations
appear depending on the magnitude of the deformation and the valence neutron configurations.
At oblate deformed region, the triangular configuration of  3$\alpha$ cluster was obtained, while 
at strong deformed prolate region, two different linear-chain configurations with valence neutrons
in $\pi$-orbit and $\sigma$-orbit were obtained. 

These cluster configurations generate clear rotational bands. The $\pi$-bond linear chain
generates a rotational band around the $\alpha$ threshold energy, while triangular and
$\sigma$-bond linear chain generate rotational bands well below and well above the threshold. The
energy of the $\pi$-bond linear chain is in reasonable agreement with the resonances observed by
the $^{4}{\rm He}+{}^{10}{\rm Be}$, while the triangular band is close to the excited states
observed by the $^9$Be($^7$Li,d)$^{14}$C reaction. The analysis of the $\alpha$ reduced width
confirms the assignment of the $\pi$-bond linear chain to the observed resonances, because the
calculated and measured widths showed reasonable agreement. Thus, the positive-parity linear-chain
formation in $^{14}{\rm C}$ looks plausible.  Furthermore, the calculation predicts that the
$\pi$-bond linear-chain will also decay to the $^{10}{\rm Be}(2^+_1)$  as well as to the
$^{10}{\rm Be}(0^+_1)$. This characteristic decay pattern will be, if measured, another evidence
of the linear-chain formation. 

In the negative-parity states, the negative-parity partners of the cluster states were also
obtained by the energy variation. However, because of the mixing with the non-cluster
configurations, these cluster configurations are fragmented into many excited states. 
As a result, many excited states that have sizable $\alpha$ reduced width are obtained, and it
makes the correspondence between the theory and experiment ambiguous. Thus, the present result is
negative for the linear-chain formation in the negative-parity, although further studies are in
need to identify the structure of the observed negative-parity resonances.

\acknowledgements
The authors acknowledges the fruitful discussions with Dr. Suhara, Dr. Kanada-En'yo, Dr. Fritsch, 
Mr. Koyama and Dr. H. Otsu. One of the authors (M.K.) acknowledges the support by the Grants-in-Aid
for Scientific Research on Innovative Areas from MEXT (Grant No. 2404:24105008) and JSPS KAKENHI
Grant No. 16K05339.

\end{document}